\documentclass[
twocolumn,
reprint,
prc,
showpacs,
a4paper,
nofootinbib,
superscriptaddress,
floatfix,
preprintnumbers,
flushbottom,
aps
]{revtex4}

\usepackage[%
	pdftitle={Elliptic flow and energy loss of heavy quarks in ultra-relativistic heavy ion collisions},%
	pdfauthor={Jan Uphoff, Oliver Fochler, Zhe Xu, and Carsten Greiner},%
	pdfsubject={Elliptic flow and energy loss of heavy quarks in ultra-relativistic heavy ion collisions},
	pdfstartview=FitH,
  pdfpagemode=UseNone,
	bookmarksopen=true
	]{hyperref}

\usepackage[
   centertags, 
   sumlimits,  
   intlimits,  
   namelimits, 
]{amsmath} %
\usepackage{amssymb}

\makeatletter
\def\tagform@#1{\maketag@@@{\ignorespaces#1\unskip\@@italiccorr}}
\let\orgtheequation\theequation
\def\theequation{(\orgtheequation)}
\makeatother

\usepackage[all]{hypcap} 
\usepackage{dcolumn}
\usepackage{multirow}
\usepackage{overpic}

\usepackage[english]{babel} 

\usepackage[%
]{graphicx}

\begin{document}

\title{Elliptic flow and energy loss of heavy quarks in ultra-relativistic heavy ion collisions}

\author{Jan Uphoff}
\email[E-mail: ]{uphoff@th.physik.uni-frankfurt.de}

\author{Oliver Fochler}
\affiliation{Institut f\"ur Theoretische Physik, Johann Wolfgang 
Goethe-Universit\"at Frankfurt, Max-von-Laue-Str. 1, 
D-60438 Frankfurt am Main, Germany}

\author{Zhe Xu}
\affiliation{Frankfurt Institute for Advanced Studies, Ruth-Moufang-Str. 1, D-60438 Frankfurt am Main, Germany}
\affiliation{Institut f\"ur Theoretische Physik, Johann Wolfgang 
Goethe-Universit\"at Frankfurt, Max-von-Laue-Str. 1, 
D-60438 Frankfurt am Main, Germany}

\author{Carsten Greiner}
\affiliation{Institut f\"ur Theoretische Physik, Johann Wolfgang 
Goethe-Universit\"at Frankfurt, Max-von-Laue-Str. 1, 
D-60438 Frankfurt am Main, Germany}

\date{\today}

\begin{abstract}
The space-time propagation of heavy quarks in ultra-relativistic heavy ion collisions is studied within the partonic transport model \emph{Boltzmann Approach of MultiParton Scatterings} (BAMPS). In this model heavy quarks interact with the partonic medium via binary scatterings. The cross sections for these interactions are calculated with leading order perturbative QCD, but feature a more precise Debye screening derived within the hard thermal loop approximation and obey the running of the coupling. Within this framework the elliptic flow and the nuclear modification factor of heavy quarks are computed for RHIC and LHC energies and compared to available experimental data. It is found that binary scatterings alone cannot reproduce the data and, therefore, radiative corrections have to be taken into account.
\end{abstract}


\maketitle

\section{Introduction}

In ultra-relativistic heavy ion collisions a medium is produced that behaves like a nearly perfect fluid \cite{Adams:2005dq,Adcox:2004mh,Arsene:2004fa,Back:2004je}, i.e. has a small viscosity to entropy density ratio. This medium is thought to be a new state of matter, which consists of quarks and gluons and, therefore, is called quark-gluon plasma (QGP).

Heavy quarks, in particular, are an ideal probe for this medium. Due to their large mass, they are produced in the early stage of the collision, where the energy density is large  \cite{Uphoff:2010sh}. Consequently, they traverse the QGP and interact with the rest of the medium. Due to these processes, their distributions are modified and can reveal -- via experimentally accessible observables like the elliptic flow and nuclear modification factor -- information about the properties of the medium. Furthermore, heavy quarks are rare and tagged by their flavor, which renders them as an unique probe, even after hadronization, due to flavor conservation.

The experimentally measured elliptic flow $v_2$ and nuclear modification factor $R_{AA}$ of heavy flavor electrons \cite{Abelev:2006db,Adare:2006nq,Adare:2010de} are comparable to that of light hadrons. This result was surprising since it was thought that the radiative part of the energy loss of heavy quarks is suppressed due to the ``dead cone effect'' \cite{Dokshitzer:2001zm,Zhang:2003wk}. Whether the large elliptic flow and strong suppression is due to collisional or radiative interactions -- or both (or even other effects) -- is under investigation \cite{Armesto:2005mz,vanHees:2005wb,Moore:2004tg,Mustafa:2004dr,Wicks:2005gt,Adil:2006ra,Peigne:2008nd,Gossiaux:2008jv,Alberico:2011zy}.

In the present study we investigate the contribution of elastic scatterings with a running coupling and an improved Debye screening inspired by hard thermal loop calculations to $v_2$ and $R_{AA}$. 

This article is organized as follows. In the next section we present our model BAMPS. Section~\ref{sec:cross_section} explains the modifications we employed to the standard leading order cross section. In Sec.~\ref{sec:v2_raa} we show our results for RHIC and LHC and compare them to the experimental data where possible. Finally, we conclude with a short summary.

\section{Parton cascade BAMPS}
\label{sec:bamps}

The \emph{Boltzmann Approach of MultiParton Scatterings} (BAMPS) \cite{Xu:2004mz,Xu:2007aa} is a $3+1$-dimensional partonic transport model, which solves the Boltzmann equation,
\begin{equation}
\label{boltzmann}
\left ( \frac{\partial}{\partial t} + \frac{{\mathbf p}_i}{E_i}
\frac{\partial}{\partial {\mathbf r}} \right )\,
f_i({\mathbf r}, {\mathbf p}_i, t) = {\cal C}_i^{2\rightarrow 2} + {\cal C}_i^{2\leftrightarrow 3}+ \ldots  \ ,
\end{equation}
for on-shell partons and perturbative QCD (pQCD) interactions. It includes elastic and also inelastic gluonic ($g$) interactions, $g g \rightarrow g g $, $
        g g g \rightarrow g g    $, and $
        g g \rightarrow g g g $, the last one being important, for instance, for thermalization \cite{Xu:2004mz,Xu:2007aa}, elliptic flow \cite{Xu:2007jv,Xu:2008av,Xu:2010cq}, or jet-quenching \cite{Fochler:2008ts,Fochler:2010wn} of gluons.

Heavy quarks ($Q$) are produced in initial hard parton scatterings or in the QGP in the reaction $g g \rightarrow Q  \bar{Q}$ (the back reaction of this process is also possible but negligible at RHIC and LHC). Studies on heavy quark production within the QGP at RHIC and LHC energies have been carried out with BAMPS and can be found in Refs.~\cite{Uphoff:2010sh,Uphoff:2010fz,Uphoff:2010sy,Uphoff:2010bv}.
In the present model heavy quarks interact with the gluonic medium via elastic collisions, $g Q \rightarrow g Q $ or $g \bar{Q} \rightarrow g \bar{Q}$. The implementation of radiative corrections such as $g Q \rightarrow g Q g $ is planned for the future.

The incorporation of light quarks ($q$) is underway and will be presented soon. However, the impact on heavy quarks will be very small, since gluons dominate the early stages of the heavy ion collision and the cross section of $q Q \rightarrow q Q $ is suppressed compared to $g Q \rightarrow g Q $ due to a smaller color factor.

The initial heavy quark distributions are generated with the Monte Carlo event generator for next-to-leading order (NLO) calculations \textsc{mc@nlo} \cite{Frixione:2002ik,Frixione:2003ei}.
To compare the initial distributions to the experimental data from proton-proton collisions, the heavy quarks are fragmented to $D$ and $B$ mesons, which, consequently, decay into heavy flavor electrons.

The Peterson fragmentation function \cite{Peterson:1982ak} of a heavy quark to a heavy meson $H$
\begin{align}
D_{H/Q} (z) = \frac{N}{z \left( 1 - \frac{1}{z} - \frac{\epsilon_Q}{ 1 - z } \right)^2}
\end{align}
is used  for the fragmentation process with $N$ being a normalization constant, $z = |\vec{p}_H|/|\vec{p}_Q|$, and $\epsilon_Q = 0.05$~($0.005$) for charm (bottom) quarks. The decay of heavy flavor mesons into electrons is performed with \textsc{pythia}~8.1 \cite{Sjostrand:2006za,Sjostrand:2007gs}.
\begin{figure}
	\centering
\includegraphics[width=1.0\linewidth]{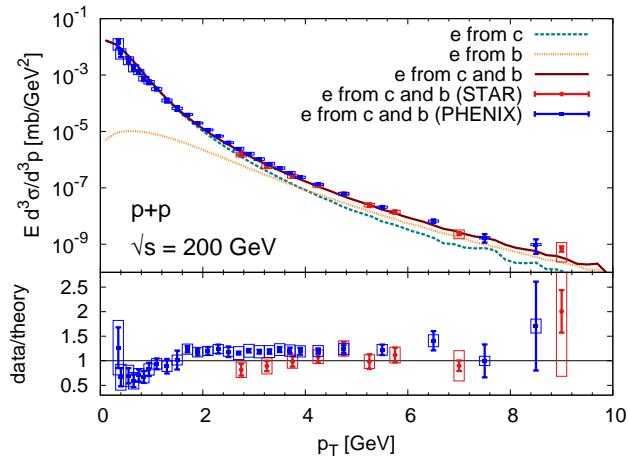}
\caption{Differential invariant cross section of heavy flavor electrons as a function of transverse momentum for proton-proton collisions with $\sqrt{s}=200 \, {\rm GeV}$ simulated with \textsc{mc@nlo} without shadowing effects. For comparison experimental data \cite{Adare:2010de,Agakishiev:2011mr} are also shown.}
\label{fig:ini_hq}
\end{figure}
Figure~\ref{fig:ini_hq} compares measured heavy flavor electrons to our initial distributions obtained with \textsc{mc@nlo}  for a factorization and renormalization scale of $\mu_F=\mu_R = 0.65\, \sqrt{p_T^2 + M_c^2}$ for charm ($M_c=1.3 \, {\rm GeV}$) and $\mu_F=\mu_R = 0.4\, \sqrt{p_T^2 + M_b^2}$ for bottom quarks ($M_b=4.6 \, {\rm GeV}$). These (in principle, arbitrary) scales are chosen in such a way that a good agreement with the experimental data is found. 

As a note, large theoretical uncertainties in the heavy quark distributions exist due to uncertainties in the parton distribution functions, renormalization and factorization scale, and heavy quark masses  \cite{Uphoff:2010sh,Cacciari:2005rk}. In particular, the relative contributions of charm and bottom quarks to the electron spectrum are not fully settled yet. Although we choose the scales such that the electron spectrum at RHIC is reproduced and employ the same parameter set for LHC initial conditions, there are considerable uncertainties in the initial heavy quark distributions at LHC.

The initial distributions for nucleus-nucleus collisions are obtained by scaling from proton-proton collision with the Glauber model. In the present study nuclear effects for the parton distribution functions, for instance, shadowing or the Cronin effect, are not taken into account for heavy quark production since the impact on the heavy quark distributions at intermediate and high transverse momentum $p_T$ is rather small \cite{Armesto:2005mz}.

The initial gluon distributions are obtained either with \textsc{pythia} or the mini-jet model. More details can be found in Refs.~\cite{Uphoff:2010sh,Xu:2004mz}.

Since charm and bottom quarks are very rare probes (there are about four heavy quark pairs at mid-rapidity in a central Au+Au collision at RHIC compared to about 800 gluons \cite{Uphoff:2010sh}), one needs to simulate several million events to yield sufficient statistics to obtain their spectra and elliptic flow. However, the most time consuming part of these event simulations is the computation of the interactions among the sizeable number of gluons, for which we do not need as many statistics. Therefore, we generate a smaller sample of pure gluonic events and use them several times as a background medium in which heavy quarks generated with \textsc{mc@nlo} are placed.
This treatment is in perfect accordance with the conventional BAMPS model. The only difference to full simulations is the neglect of medium modifications induced by the heavy quarks, which is a very good approximation given the small number of heavy quarks in a heavy ion collision.

\section{Modification of the cross section}
\label{sec:cross_section}

The cross sections for $Q + g \rightarrow Q + g$ can be calculated in leading-order (LO) pQCD. To treat this accurately, we explicitly take the running of the coupling into account. Furthermore, we employ a more precise Debye screening compared to the standard procedure. This is done by comparing the energy loss calculated with the LO cross section and the hard thermal loop (HTL) approximation \cite{Gossiaux:2008jv,Peshier:2008bg}.

The LO cross section \cite{Combridge:1978kx} diverges for small Mandelstam $t$ due to the gluon propagator in the $t$ channel. In thermal field theory, however, long-range interactions (which correspond to small $t$) are screened by the medium, an effect that originates formally from thermal loop corrections. 
The resummation and renormalization of these loop corrections results in a screening of the gluon propagator with its thermal self-energy $\Pi_{\rm therm}(\omega,q)$ and replacing the bare coupling with the running coupling $\alpha_s(t)$ \cite{Peshier:2006hi,Peshier:2008bg}
\begin{align}
\label{t_channel_screening_selfe}
	 \frac{\alpha_s}{t} \rightarrow \frac{\alpha_s(t)}{t-\Pi_{\rm therm}(\omega,q)} \ ,
\end{align}
with $(\omega, {\bf q} ) = P_1^\mu-P_3^\mu$ (the four-momentum difference of the incoming and outgoing heavy quark in the center-of-mass frame). Nevertheless, calculations with the self-energy are too involved to be incorporated in a transport model. However, we approximate the self-energy by a screening mass $\mu^2 = \kappa \, m_{D}^2$, which is proportional to the Debye mass $m_{D}^2$,
\begin{align}
\label{t_channel_screening_debye}
	 \frac{\alpha_s}{t} \rightarrow \frac{\alpha_s(t)}{t-\mu^2} = \frac{\alpha_s(t)}{t-\kappa \, m_{D}^2} \ .
\end{align}

The Debye mass is calculated by
\cite{Xu:2004mz}  
\begin{equation}
\label{md2}
m_D^2 = \pi \alpha_s \nu_g \int \frac{{\rm d}^3p}{(2\pi)^3} \frac{1}{p} 
( N_c f_g + n_f f_q) \ ,
\end{equation}
where $N_c=3$ denotes the number of colors, $\nu_g = 16$ the gluon degeneracy and $f$ the distributions of gluons and quarks. In equilibrium and with Boltzmann statistics this simplifies to $m_{D,{\rm eq}}^2 = \frac{8 \alpha_s}{\pi} (N_c+n_f) \, T^2$. For consistency we employ a running $\alpha_s = \alpha_s(t)$ also in the calculation of the Debye mass.

The prefactor $\kappa$ in Eq.~\ref{t_channel_screening_debye} is mostly set to 1 in the literature without a sophisticated reason. However, one can fix this factor analytically by comparing the energy loss per unit length ${\rm d}E/{\rm d}x$ of the LO cross section including $\kappa$ (see Eq.~\ref{t_channel_screening_debye}) to the energy loss within the hard thermal loop approach \cite{Gossiaux:2008jv,Peshier:2008bg}.

In the following we will outline the calculation for a constant coupling $\alpha_s$ since, in this case, $\kappa$ can be determined analytically. The generalization to a running coupling is given in Ref.~\cite{Peigne:2008nd}.
The heavy quark collisional energy loss within the HTL approximation was calculated in Refs.~\cite{Braaten:1991we,Peigne:2007sd,Peigne:2008nd} for quantum statistics. Analogously, for Boltzmann statistics\footnote{BAMPS treats all particles as Boltzmann particles.} the collisional energy loss of a heavy quark with energy $E$ and mass $M$ in a thermalized medium with temperature $T$ is given in the high energy limit by 
\begin{align}
\label{dedx_ana}
	\frac{{\rm d} E}{{\rm d} x}  = \frac{8  \alpha_s^2 T^2}{\pi}  
\left[ \left( 1 +\frac{n_f}{3} \right) \ln\frac{E T}{m_D^2} + \frac{2}{9} \ln\frac{E T}{M^2} 
+ f(n_f) \right] \ ,
\end{align}
with
\begin{align}
\label{dedx_fnf}
	f(n_f) = g\, n_f +h \simeq 0.251 \, n_f + 0.747   \, ,
\end{align}
where $g = \ln{2} -1/4 -\gamma/3$ and $h= (31/9) \ln{2} -101/108 -11\gamma /9$ with $\gamma$ being the Euler-Mascheroni constant and $n_f$ the number of flavors.

The energy loss calculation above is done within the HTL approach. An analogous calculation
can be repeated with the screened (according to Eq.~\ref{t_channel_screening_debye}) LO cross section instead of the full HTL calculation. One ends up with a result which is  very similar to Eq.~\ref{dedx_ana}:
\begin{align}
	\frac{{\rm d} E}{{\rm d} x}  = \frac{8  \alpha_s^2 T^2}{\pi}  
\left[ \left( 1 +\frac{n_f}{3} \right) \ln{\frac{E T}{ 2e  \kappa \, m_D^2}} + \frac{2}{9} \ln\frac{E T}{M^2} 
+ f(n_f) \right] \ .
\end{align}
Only the argument of the logarithm in the first term changes. From this, one can read off, that the screening prefactor $\kappa$ has to be $\kappa = 1/(2e) \approx 0.184 \approx 0.2$ to obtain the same result as in the HTL approximation \cite{Gossiaux:2008jv,Peshier:2008bg}. The determination of $\kappa$ for a running coupling yields also a result close to this value \cite{Gossiaux:2008jv}.

\begin{figure}
	\centering
\includegraphics[width=1.0\linewidth]{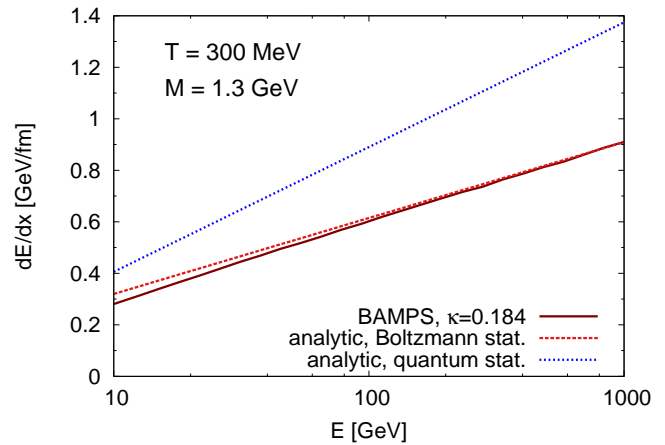}
\caption{Energy loss per unit length ${\rm d}E/{\rm d}x$ of a heavy quark jet in a static and thermalized medium of gluons ($n_f = 0$) with temperature $T = 300\, {\rm MeV}$ as a function of the jet energy $E$. The numerical result from BAMPS with the LO cross section, a constant coupling $\alpha_s = 0.3$, and a screening mass with $\kappa =  0.184$ is compared to the analytical result from Eq.~\ref{dedx_ana} for Boltzmann statistics. In addition, the curve for quantum statistics \cite{Peigne:2007sd,Peigne:2008nd} is also shown.}
\label{fig:dedx_ana_bamps}
\end{figure}

In Fig.~\ref{fig:dedx_ana_bamps} the numerical result with BAMPS and $\kappa =  0.184$ is compared to the analytic formula \eqref{dedx_ana}.
For large jet energies, that is, the regime in which the analytic formula is valid, the agreement with the numerical result is very good. For comparison the analytic curve for quantum statistics is shown as well, which is about 25\,\% larger than the Boltzmann curve for $E\simeq 10\, {\rm GeV}$.

The running coupling is evaluated at the renormalization scale $\mu_R$. While complete results are $\mu_R$ independent due to renormalization group flow equations, approximate results do have a residual $\mu_R$ dependence. The resulting uncertainty can be reduced if one chooses a scale relevant for the physical problem at hand. In the $t$ channel, for instance, large logarithms of $t$ occur in next-to-leading order due to vacuum contributions to the self-energy and vertex corrections. These  logarithms can be absorbed in $\alpha_s$ via renormalization by  choosing $\mu_R^2 = t$ \cite{Peshier:2006hi,Peigne:2008nd}. An analogous line of argument holds also for the $s$ and $u$ channels using their characteristic momenta instead of $t$ \cite{Peshier:2008bg,Peigne:2008nd}.
Thus, we evaluate $\alpha_s(\mu_R^2)$ at the virtuality of the respective channel, that is, $s-M^2$, $t$, and $u-M^2$ for the $s$, $t$, and $u$ channels, respectively. 

An effective description of the running coupling can be obtained from measurements of $e^+e^-$ annihilation and non-strange hadronic decays of $\tau$ leptons and continued to the time-like region \cite{Dokshitzer:1995qm,Gossiaux:2008jv,Peshier:2008bg}: 
\begin{align}
\label{alpha_s_continued}
 \alpha_s(Q^2)= \frac{4\pi}{\beta_0} \begin{cases}
  L_-^{-1}  & Q^2 < 0\\
  \frac12 - \pi^{-1} {\rm arctan}( L_+/\pi ) &  Q^2 > 0
\end{cases}
\end{align}
with $\beta_0 = 11-\frac23\, n_f$  and
$L_\pm = \ln(\pm Q^2/\Lambda^2)$ with $\Lambda=200 \, {\rm MeV}$. If $\alpha_s(Q^2)$ is larger than $\alpha_s^{\rm max} = 1.0$ it is set to $\alpha_s^{\rm max}$. 
This cutoff procedure is in line with a soft average of $\alpha_s$ given by the universality hypothesis \cite{Peshier:2008bg,Dokshitzer:2003qj}. 
We checked that our results of the energy loss are not very sensitive on the exact value of $\Lambda$ or $\alpha_s^{\rm max}$ since the contribution from the soft part of the process is very small.

\section{Elliptic flow and nuclear modification factor}
\label{sec:v2_raa}

The elliptic flow 
\begin{align}
\label{elliptic_flow}
  v_2=\left\langle  \frac{p_x^2 -p_y^2}{p_T^2}\right\rangle 
\end{align} 
($p_x$ and $p_y$ are the momenta in the $x$ and $y$ directions with respect to the reaction plane)
and the nuclear modification factor
\begin{align}
  R_{AA}=\frac{{\rm d}^{2}N_{AA}/{\rm d}p_{T}{\rm d}y}{N_{\rm bin} \, {\rm d}^{2}N_{pp}/{\rm d}p_{T}{\rm d}y}
\end{align} 
are suitable observables to study the energy loss of heavy quarks and their coupling to the medium. The larger the $v_2$ the stronger heavy quarks interact with the medium and adopt its momentum anisotropy. A small value of $R_{AA}$ indicates a strong suppression and, therefore, large energy loss of heavy quarks in the medium. Unfortunately, $D$ and $B$ mesons that stem from charm and bottom quarks, respectively, cannot be reconstructed directly at RHIC yet. Therefore, one measures single electrons that are decay products of $D$ and $B$ mesons. The $v_2$ and $R_{AA}$ of these heavy flavor electrons are comparable to the respective values for light hadrons \cite{Abelev:2006db,Adare:2006nq,Adare:2010de}.

To compare to the experimental data we also perform the fragmentation of heavy quarks to $D$ and $B$ mesons and the consecutive decay to electrons. The final distribution of the electrons is very similar to that of heavy quarks, only shifted to lower transverse momentum $p_T$. For small $p_T$ Peterson fragmentation is not a good description of the hadronization process and other schemes like coalescence might be more appropriate. However, we postpone the implementation of more sophisticated hadronization schemes in BAMPS to the future.

For elastic heavy quark scatterings with the medium, the LO pQCD cross section with a constant coupling and a standard Debye screening is too small to explain the measured data \cite{Uphoff:2010fz}. However, the improvements presented in Sec.~\ref{sec:cross_section} increase the cross section and yield results which are much closer to the experimental data.

\begin{figure}


%

\centering
\begin{overpic}[width=1.0\linewidth]{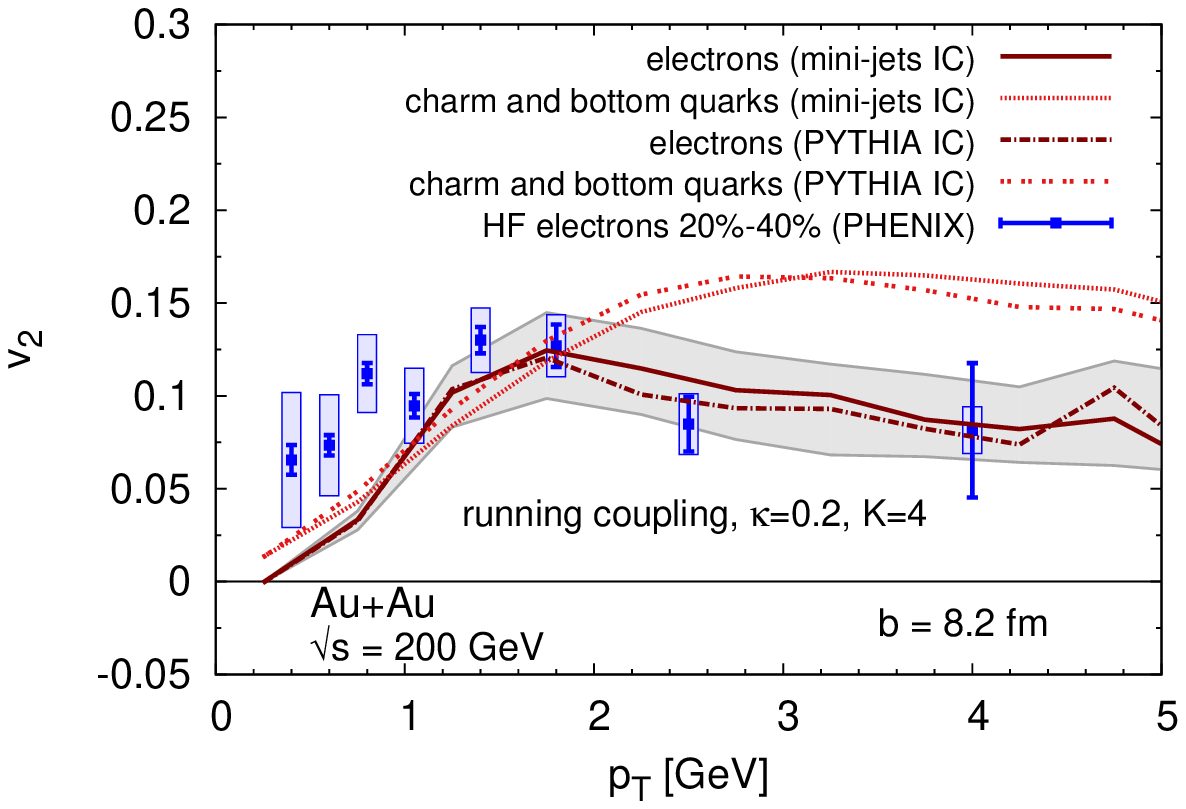}
\put(22,60){(a)} 
\end{overpic}

\begin{overpic}[width=1.0\linewidth]{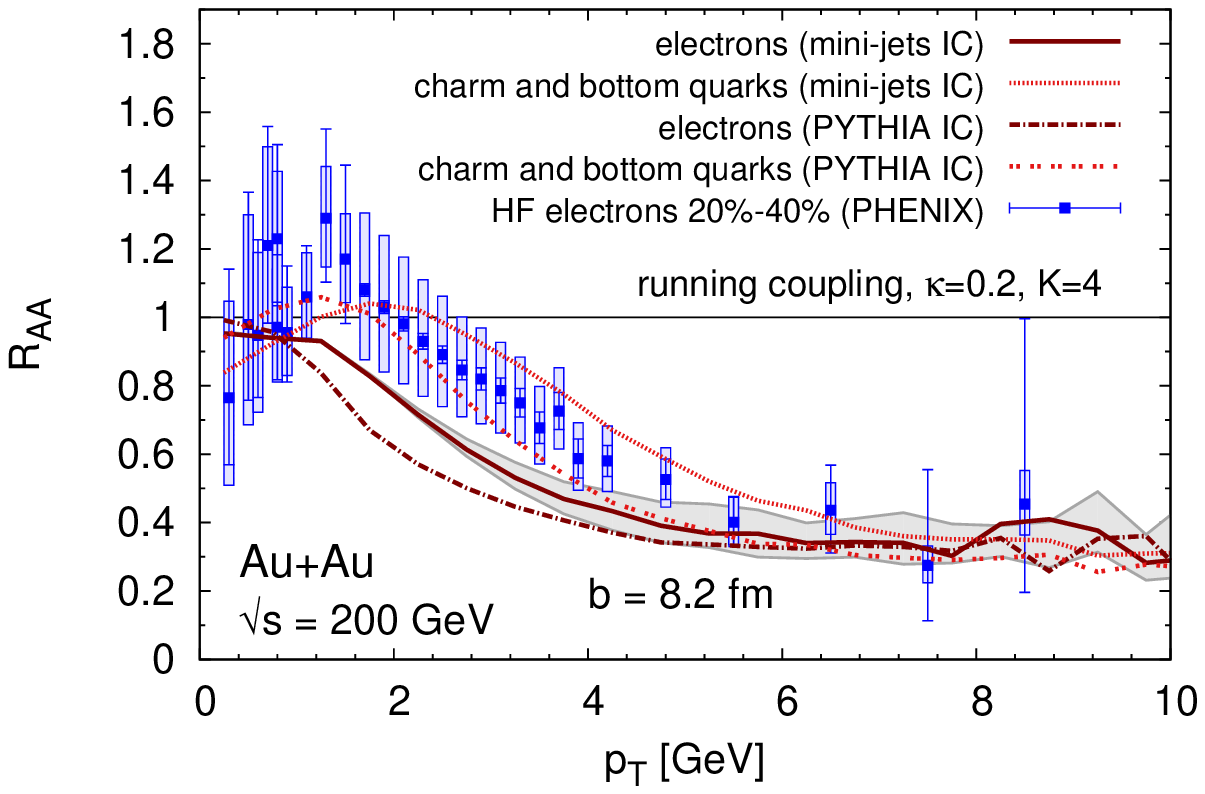}
\put(22,60){(b)} 
\end{overpic}

\caption{(a) Elliptic flow $v_2$ and (b) nuclear modification factor $R_{AA}$ of heavy quarks and heavy flavor electrons with pseudo-rapidity $|\eta|<0.35$ for Au+Au collisions at RHIC with an impact parameter of $b=8.2 \, {\rm fm}$. The curves are obtained with \textsc{pythia} and mini-jet initial conditions (IC) for the gluons. The cross section of $gQ \rightarrow gQ$ is multiplied with the factor $K=4$. To estimate the uncertainty of this $K$ factor we plotted for mini-jet IC the electron curves for $K=3$ and $K=5$ as gray bands. For comparison, data of heavy flavor electrons for the centrality class of 20\,\%$-$40\,\% \cite{Adare:2010de} are shown.}
\label{fig:v2_raa_rhic}
\end{figure}

In Fig.~\ref{fig:v2_raa_rhic} we show the elliptic flow and nuclear modification factor of heavy quarks and heavy flavor electrons for a non-central Au+Au collision at RHIC obtained with BAMPS at the end of the QGP phase, that is, after the energy density of a given cell has dropped below $0.6 \, {\rm GeV/fm^3}$ and interactions are not allowed any more \cite{Xu:2008av}.
For these curves the cross section of $gQ \rightarrow gQ$ has been multiplied with an artificial factor $K=4$ to be compatible with the data. 
BAMPS studies on the energy loss of light partons reveal that radiative contributions are dominant for light particles \cite{Fochler:2008ts,Fochler:2010wn}. For heavy quarks these contributions are expected to be suppressed due to the dead cone effect, but it has been shown that they are still slightly larger than the elastic contributions \cite{Mustafa:2004dr,Wicks:2005gt}. 
Consequently, we assume that the implementation of radiative corrections like $gQ \rightarrow gQg$ could account for the missing factor of 4. However, it needs to and will be checked in a forthcoming study whether these contributions have indeed the same effect as a constant $K$ factor. The implementation of NLO processes for heavy quarks will complement $2 \leftrightarrow 3$ interactions for gluons, which are already present in BAMPS \cite{Xu:2004mz}. Furthermore, the consideration of quantum statistics would also enhance the cross section as is shown in Fig.~\ref{fig:dedx_ana_bamps} for the energy loss and can, therefore, explain part of the missing factor of 4.

Of course, a constant factor $K=4$ is only an estimate for radiative contributions and quantum statistics. To explore the uncertainty of this factor the curves for $K=3$ and 5 are also plotted in Fig.~\ref{fig:v2_raa_rhic} as gray bands for electrons and mini-jet initial conditions. Both observables $v_2$ and $R_{AA}$ are not very sensitive on the exact value of $K$, although $K=4$ gives the best agreement with the data.

Figure~\ref{fig:v2_raa_rhic} compares the elliptic flow and heavy quark suppression obtained with two different initial conditions for the gluons. Both the \textsc{pythia} and the mini-jet scenario lead to comparable curves, especially for larger $p_T$. This indicates that both heavy flavor observables are not very sensitive on the initial light parton distributions as long as the elliptic flow of light hadrons is reproduced, which has been shown for mini-jet initial conditions in Refs.~\cite{Xu:2007jv,Xu:2008av,Xu:2010cq}. In addition, we confirmed this also with \textsc{pythia} initial conditions and will present the results in a forthcoming paper. However, in Ref.~\cite{Gossiaux:2011ea} it was found that differences in the medium evolution can lead to modifications of the heavy quark suppression and elliptic flow up to a factor of 2.

With $K=4$ the agreement with the data is very good for large $p_T$. We emphasize that both $v_2$ and $R_{AA}$ are described simultaneously within the same partonic transport model. For small $p_T$ Peterson fragmentation is not suitable and coalescence might play an important role, which modifies the $v_2$ and $R_{AA}$ due to the contribution of light quarks.

These results are obtained with initial heavy quark distributions from \textsc{mc@nlo}. In addition, we employed heavy quark initial conditions from NLO calculations of Refs.~\cite{Cacciari:2005rk,Vogt2010} and found a good agreement with the curves shown in Fig.~\ref{fig:v2_raa_rhic}. In contrast to \textsc{pythia} which we used in Refs.~\cite{Uphoff:2010sy,Uphoff:2010bv} for the initial heavy quark distributions, the electron spectrum from \textsc{mc@nlo} is slightly steeper for low $p_T$ which results in a less suppressed $R_{AA}$ and, therefore, better agreement with the experimental data for $p_T < 3\, \rm{GeV}$.

In previous studies \cite{Uphoff:2010sh,Uphoff:2010fz,Uphoff:2010bv} we have found that at LHC energies a sizeable fraction of the produced charm quarks is created during the evolution of the QGP and not only in the initial hard parton scatterings. These secondarily produced charm quarks can, of course, contribute to the $v_2$ and $R_{AA}$ of heavy quarks. 
However, the impact on these observables is very small and only affects the region of low $p_T$ since secondary charm quarks are produced with small momenta. Due to their early production time which is usually less than $1 \, {\rm fm}/c$ \cite{Uphoff:2010sh} secondary charm quarks have enough time to interact with the medium, lose energy, and build up elliptic flow. Consequently, also in the low  $p_T$ region, $v_2$ is barely changed by secondary charm quarks.
The fraction of bottom quarks produced in the medium is so small that the assumption that all of them are created in the initial hard parton scatterings is justified \cite{Uphoff:2010sh}.
In short, since the contribution of secondarily produced heavy quarks is insignificant we neglect it in the following. 

In addition to the uncertainties of the initial heavy quark distributions (see Sec.~\ref{sec:bamps}), there are also sizeable uncertainties concerning the bulk medium at LHC. For initial gluon conditions from \textsc{pythia} we obtain a final ${\rm d}N^g/{\rm d}y \approx 1770$ and ${\rm d}E_T^g/{\rm d}y \approx 1570 \, \rm{GeV}$ of gluons at mid-rapidity in central collisions after the energy density has dropped below $0.6 \, {\rm GeV/fm^3}$ for every cell. Unfortunately, experimental data for ${\rm d}E_T/{\rm d}y$ is not available yet and it is rather involved to compare the measured ${\rm d}N_{\rm ch}/{\rm d}\eta$ of charged hadrons to the number of gluons due to non-perturbative effects at hadronization. However, we checked that the ratio of ${\rm d}N_{\rm ch}/{\rm d}\eta$ at LHC \cite{Aamodt:2010cz} and ${\rm d}N_{\rm ch}/{\rm d}\eta$ at RHIC \cite{Adler:2004zn} for central collisions, (without taking errors into account)
\begin{align}
	R_{\rm ch} := \frac{ {\rm d}N_{\rm ch}/{\rm d}\eta|_{\rm LHC} }{ {\rm d}N_{\rm ch}/{\rm d}\eta|_{\rm RHIC} } = \frac{ 1601 }{ 687 } \approx 2.33 \ ,
\end{align}
is nearly equal to the ratio of the final ${\rm d}N^g/{\rm d}y$ at LHC and RHIC obtained with BAMPS, which is for \textsc{pythia} initial conditions
\begin{align}
	R_{g} := \frac{ {\rm d}N^g/{\rm d}y|_{\rm LHC} }{ {\rm d}N^g/{\rm d}y|_{\rm RHIC} } \approx \frac{ 1770 }{ 740 } \approx 2.39 \ .
\end{align}
This indicates that the scaling to LHC has been performed accurately. Nevertheless, a comparison to measured ${\rm d}E_T/{\rm d}y$ will be an additional test.

For mini-jet initial conditions one can vary the momentum cutoff $p_0$ for Pb+Pb collisions at the LHC to obtain an $R_{g}$ which is close to $R_{\rm ch}$. This procedure leads to $p_0 = 3.5 \, \rm{GeV}$ and
\begin{align}
	R_{g} \approx \frac{ 1820 }{ 800 } \approx 2.28 \ .
\end{align}
The final transverse energy distribution for this cutoff is ${\rm d}E_T^g/{\rm d}y \approx 1810 \, \rm{GeV}$.

For an impact parameter of $b=8.2 \, {\rm fm}$ the final values of gluons at mid-rapidity are ${\rm d}N^g/{\rm d}y \approx 450 \, (440)$ and ${\rm d}E_T^g/{\rm d}y \approx 410 \, \rm{GeV} \, (570 \, \rm{GeV})$ for \textsc{pythia} (mini-jet with $p_0 = 3.5 \, \rm{GeV}$) initial conditions. The ratio of the former to our RHIC value of ${\rm d}N^g/{\rm d}y \approx 210 \, (200)$ for $b=8.2 \, {\rm fm}$ and \textsc{pythia} (mini-jet) initial conditions is $R_{g}=2.14 \, (2.20)$ which is close to the experimental ratio of LHC and RHIC of $R_{\rm ch} = 2.37$ for the centrality class 20\,\%$-$30\,\% \cite{Aamodt:2010cz,Adler:2004zn}.

\begin{figure*}

\centering
\begin{minipage}[t]{0.49\textwidth}
\centering
\begin{overpic}[width=1.0\linewidth]{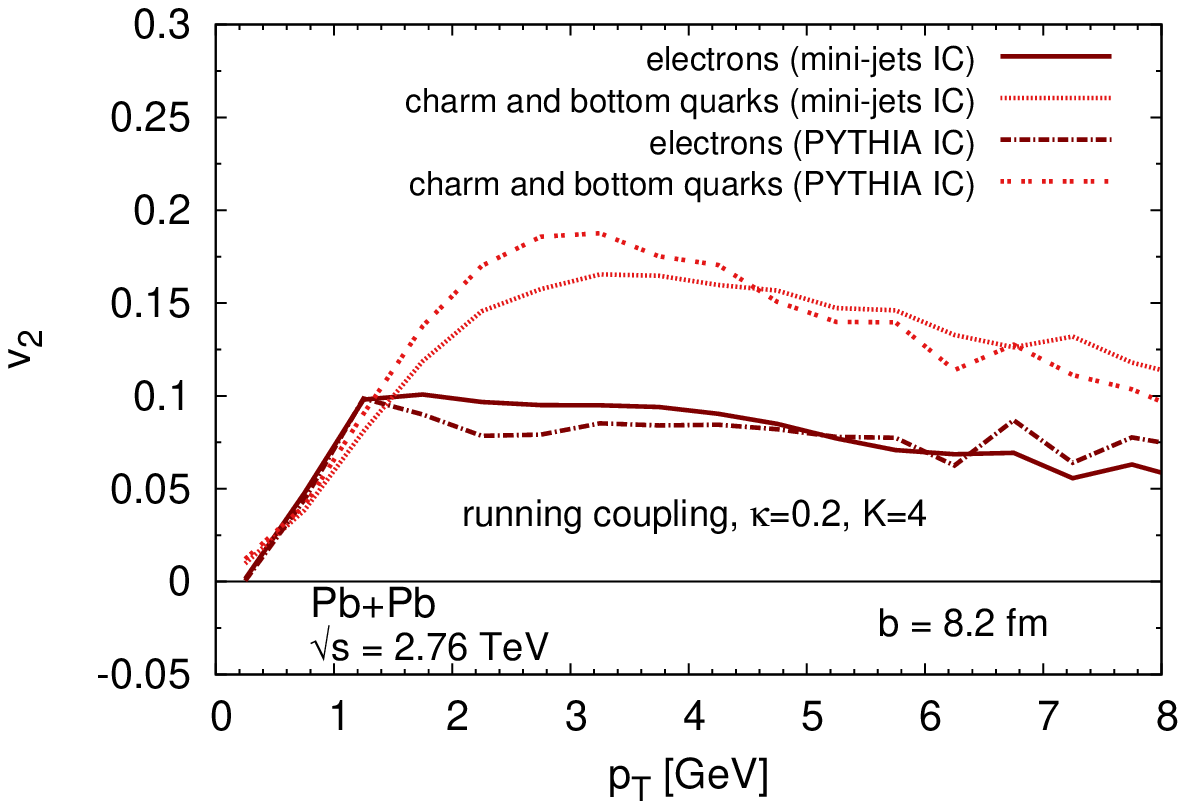}
\put(22,60){(a)}
\end{overpic}
\end{minipage}
\hfill
\begin{minipage}[t]{0.49\textwidth}
\centering
\begin{overpic}[width=1.0\linewidth]{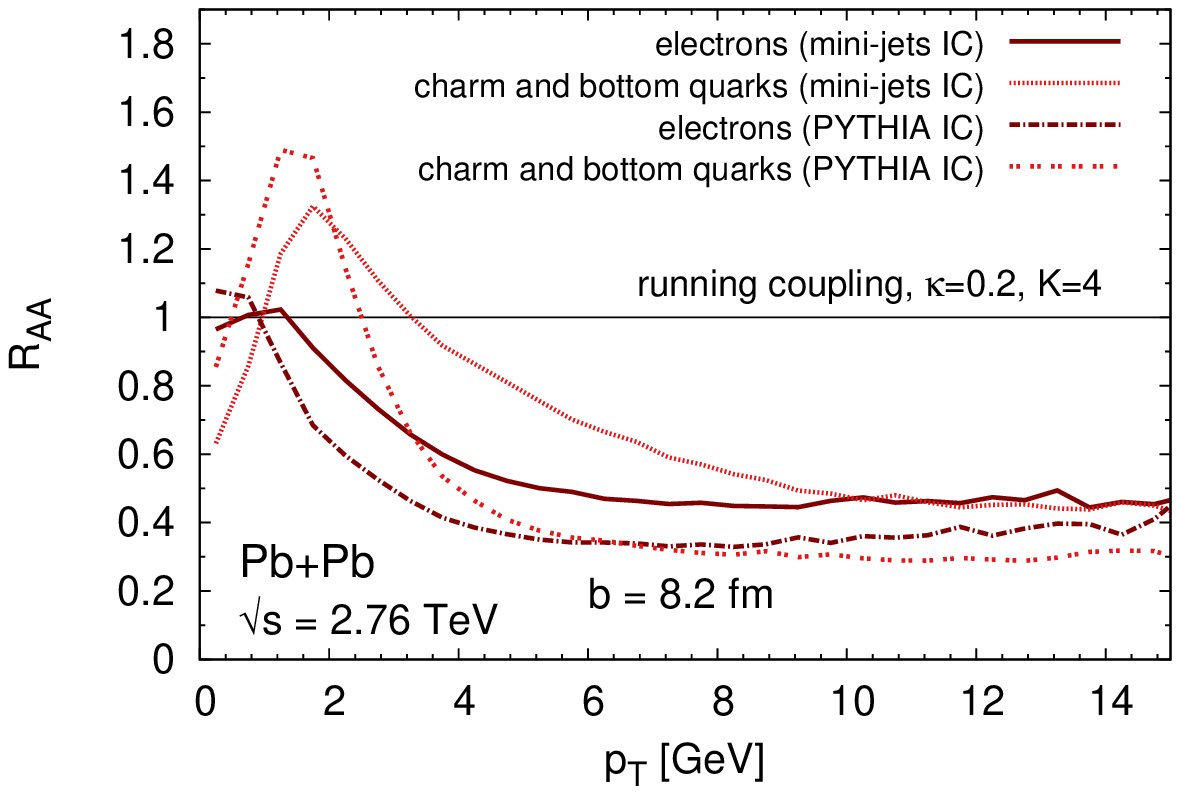}
\put(22,60){(b)} 
\end{overpic}
\end{minipage}

%

\caption{As in Fig.~\ref{fig:v2_raa_rhic}, but for Pb+Pb collisions at LHC with $\sqrt{s_{NN}}=2.76 \, \rm{TeV}$. 
}
\label{fig:v2_raa_lhc}
\end{figure*}

Figure~\ref{fig:v2_raa_lhc} shows $v_2$ and $R_{AA}$ of initially produced heavy quarks for a Pb+Pb collision at LHC with $\sqrt{s_{NN}}=2.76 \, \rm{TeV}$ and an impact parameter of $b=8.2 \, {\rm fm}$.
As for RHIC (cf. Fig.~\ref{fig:v2_raa_rhic}) we employ a cross section for the heavy quark interactions with gluons, which incorporates the running coupling, the improved Debye screening, and $K=4$.
Both for heavy quarks and heavy flavor electrons the elliptic flow for \textsc{pythia} and mini-jet initial conditions agree very well. Heavy quarks and thus also heavy flavor electrons are slightly less suppressed for mini-jet initial conditions compared to \textsc{pythia} initial conditions.
This effect can also be seen for gluons and is probably due to a smaller initial gluon number at mid-rapidity for the mini-jet case compared to the \textsc{pythia} scenario, which results in a less opaque medium before the QGP is chemically equilibrated.

The elliptic flow and suppression of electrons from charm quarks are slightly larger than at RHIC. However, the contribution from bottom quarks becomes important also at smaller $p_T$. As a consequence, the $v_2$ and $R_{AA}$ at LHC are very similar to the RHIC results.

\section{Conclusions}

We presented the BAMPS results of the elliptic flow and nuclear modification factor of heavy flavor electrons at RHIC and LHC. For this study, elastic interactions with the partonic medium are taken into account in LO pQCD. The cross section of these processes obey the running of the coupling and include a Debye screening motivated from HTL calculations, which is more precise compared to previous approaches. However, elastic scattering alone does not lead to the sizeable elliptic flow and nuclear suppression measured at RHIC. To yield the same values as the data, the cross section must be multiplied with a factor $K=4$. We assume that radiative contributions to the interaction can account for this phenomenological factor, which we will check in an upcoming study.
The elliptic flow and nuclear modification factor at LHC are found to be of the same order as at RHIC.

\section*{Acknowledgements}
J.U. would like to thank A. Peshier for stimulating and helpful discussions and the kind hospitality at the University of Cape Town, where part of this work has been done.

The BAMPS simulations were performed at the Center for Scientific Computing of the Goethe University Frankfurt. This work was supported by the Helmholtz International Center for FAIR within the framework of the LOEWE program launched by the State of Hesse.

\bibliography{hq}

\end{document}